\documentclass[aps,amsmath,amsfonts,11pt,nofootinbib]{revtex4}
\begin{document}

\def\er#1{eqn.\eqref{#1}}
\def\nn{\nonumber}
\setlength\arraycolsep{2pt}  
\newcommand{\td}{\textup{d}} 

\def\bmfactor{\big(b-\frac{{x^i}^2}{4}\big)}
\title{NS5 Brane and Little String Duality in the pp-wave Limit}
\author{P.~Matlock}
\email{pwm@sfu.ca}
\affiliation{Department of Physics, Simon Fraser University, Burnaby BC, Canada}
\author{R.~Parthasarathy}
\email{sarathy@imsc.res.in}
\affiliation{The Institute of Mathematical Sciences, Chennai - 600 113, India}
\author{K.~S.~Viswanathan}
\email{kviswana@sfu.ca}
\affiliation{Department of Physics, Simon Fraser University, Burnaby BC, Canada}
\author{Y.~Yang}
\email{yyangc@sfu.ca}
\affiliation{Department of Physics, Simon Fraser University, Burnaby BC, Canada}

\begin{abstract}

  We study NSR strings in the Nappi-Witten background, which is the
Penrose limit of a certain NS5-brane supergravity solution.  We solve
the theory in the light-cone gauge, obtaining the spectrum, which is
space-time supersymmetric.  In light of the LST/NS5-brane duality,
this spectrum should be in correspondence with the states of little
string theory in the appropriate limit. A semiclassical analysis
verifies that the relationship between energy and angular momentum, after a
field redefinition, matches the known result for a flat background.
\end{abstract}

\maketitle

\section{Introduction}

Of recent interest have been consistent six-dimensional non-local theories
which do not contain gravity, but which exhibit stringy properties such as T-duality
and a Hagedorn density of states \cite{0010169}. These so-called Little String Theories (LSTs) 
\cite{9707250,9705221,9909110,0301059}
are thus interesting examples of theories which are `in-between' field theories
and string theories, and are expected to shed light on both the interpretation
of non-local field theory and string theory. A recent and very complete review of all 
the main points of little string theories has been given in \cite{9911147}. 
We here mention a few basic properties; for further detail the reader is referred
to that review and references therein.

Generically, little string theory is defined by first considering some background of NS5-branes in
type-II string theory. Taking a limit in which the string coupling $g_s$ goes
to zero while the string mass $M_s$ is held fixed leads to a free theory in the bulk,
decoupled from a theory living on each NS5-brane; this is the LST \cite{9804042,9904142}. The string scale 
$M_s=1/l_s$ is the only defining parameter for the LST, and is important in the following way.
NS5-branes are obtained from M-theory 5-branes by compactifying on a transverse circle $S^1$.
The resultant string theory with NS5-branes exhibits T-duality with respect to this circle,
and an NS5-brane wrapping the circle of radius $R$ will map in the dual theory to an NS5-brane
wrapping a circle of radius $1/R{M_s^2}$. Now, from either of these theories can be defined
a little string theory by taking the limit mentioned above, and the LST will inherit the 
T-duality from the string theory. This is an indication that the resulting theory will be
non-local on the scale of $M_s$, and it is somewhat surprising that a non-gravitational
theory can exhibit such T-duality.

In type-II theories, NS5-brane solutions break half of the supersymmetry.
Ten-dimensional supersymmetry, dimensionally reduced to a six dimensional world-volume, 
will result in a chiral theory with $(2,2)$ supersymmetry. The NS5-brane solution may
break this in either of two ways, resulting in either $(2,0)$ or $(1,1)$ supersymmetry 
for type IIA or IIB string theory, respectively. Considering $N$ NS5-branes,
at low energies LST must reduce to a superconformal $(2,0)$ theory on $N$ M5-branes for 
the first case, or a $(1,1)$ $U(N)$ gauge theory in the second, with $G_{\textup{YM}}=1/M_s$.

LST has a holographic description \cite{9808149}, which turns out to be useful for explicit computations
of, for instance, the spectrum and the density of states.
This description involves first considering a background of M5-branes, in the so-called
linear dilaton form,\cite{9808149} in which $g_s$ goes as the exponential of some coordinate $\phi$.
$\phi\rightarrow\infty$ is then the `boundary,' where the limit $g_s\rightarrow0$
obtains, and a holographic LST may be constructed in this way.
In particular, the string theory may be used to calculate the spectrum of states in the LST,
and, for example, correlation functions in the $(2,0)$ LST may be obtained from
supergravity.

Although many theories are expected to be holographic, the AdS/CFT correspondence 
is still the most tested and trusted example. String theory on $AdS_5\times S^5$ is
dual to $d=4$, $\mathcal{N}=4$ SYM theory. Taking the Penrose limit of the `left-hand side'
of this duality is equivalent to selecting a specific sector of large-dimension and
large-charge operators in the SYM theory \cite{0202021}. In the present case,
string theory on an NS5-brane background corresponds to LST on the NS5-brane worldvolume.
Generally it is not possible to solve string theory exactly in an NS5-brane 
background, but solubility obtains in the pp-wave limit.
It was first identified in \cite{0202157} that taking a Penrose limit in this 
case will reduce the NS5-brane metric to the Nappi-Witten form \cite{0202179,9310112,9805006,9909164},
and this in turn corresponds to the high-energy sector of the LST \cite{0205258,0303212}.

With this correspondence in mind, we wish in the present paper to consider type-II
string theory in the Nappi-Witten background. The bosonic case has been analysed in detail
in \cite{9503222}, and we here treat the full supersymmetric theory. In light-cone gauge,
the theory is found to be completely soluble, and the Hamiltonian is calculated, giving
the spectrum. The high-energy sector of the LST corresponding to the parent theory on 
the NS5-brane metric is expected to share this spectrum. In particular, this means that 
at least this sector of the LST must be supersymmetric.

In the following section, we summarise some aspects of the NS5-brane geometry of interest
and its pp-wave limit. In section \ref{smas} we perform the above computations and obtain the
spectrum. In section \ref{disc}, we compare the resulting energy-spin relation with that 
obtained by semi-classical methods along the lines of \cite{0204226}.

\section{Type IIA NS5-brane background and pp-wave limit}

In this section, we briefly review some important aspects of NS5-brane background geometries.
For further detail, in the context of little string theory, the reader is referred to the 
comprehensive review \cite{9911147}.
As we mentioned, the type IIA NS5-brane background may be described as an M-theory 5-brane
transversely compactified on $S^{1}$. From \cite{9802042,9808149}, we see that the supergravity
background corresponding to $N$ coincident extremal M5-branes is
\begin{equation}
ds^{2}=H^{-1/3}\left[dx_{6}^{2}+H\left(dx_{11}^{2}+dr^{2}+r^{2}d\Omega_{3}^{2}\right)\right]
\end{equation}
with 
\begin{equation}
H=1+\sum_{n=-\infty }^{\infty }\frac{Nl_{p}^{3}}{\left[ r^{2}+\left(
x_{11}-nR_{11}\right) ^{2}\right] ^{3/2}}
,\end{equation}
where the 11${}^\textup{th}$ dimension is the compact $S^1$ with radius $R_{11}$, and $l_p$ is
the 11-dimensional Planck length. Defining new scaled coordinates by
\begin{equation}
U=r/l_{p}^{3},\text{ \ \ }y_{11}=x_{11}/l_{p}^{3},\text{ \ \ }
R_{11}=l_{p}^{3}/l_{s}^{2},
\end{equation}
the metric becomes 
\begin{equation}
ds^{2}=l_{p}^{2}\tilde{H}^{-1/3}\left[ dx_{6}^{2}+\tilde{H}\left(
dy_{11}^{2}+dU^{2}+U^{2}d\Omega _{3}^{2}\right) \right]
\label{11d metric}
\end{equation}
with 
\begin{equation}
\tilde{H}=l_p^6+\sum_{n=-\infty }^{\infty }\frac{N}{\left[U^{2}+\left(
y_{11}-n/l_{s}^{2}\right) ^{2}\right] ^{3/2}},  \label{H}
\end{equation}
where $l_{s}\sim \frac{1}{m_{s}}$ is the `coupling constant' for the
theory on the brane, which is decoupled from the bulk theory.

For $U\ll \frac{1}{l_{s}^{2}}$,
the summation in \er{H} may be approximated by the $n=0$ term;
in this case the metric \eqref{11d metric} assumes the form of $AdS_{7}\times S^{4}$.
on which string theory is dual to the six-dimensional (2,0) SCFT. \cite{9808149}

Conversely, for $U\gg \frac{1}{l_{s}^{2}}$, the summation in \er{H}
effectively becomes an integration \cite{9808149,9904142}. The metric of $N$ coincident
extremal NS5-branes is then obtained; (up to a conformal factor and without the $dy_{11}^{2}$ term)
\begin{eqnarray}
ds_{str}^{2} &=&dx_{6}^{2}+A\left( U\right) \left( dU^{2}+U^{2}d\Omega
_{3}^{2}\right) ,  \label{NS5} \\
e^{2\Phi } &=&g_{s}^{2}A\left( U\right)
\end{eqnarray}
with 
\begin{equation}
A\left( U\right) =l_p^6+\frac{Nl_{s}^{2}}{U^{2}}.
\end{equation}
We see that the energy scale $\sqrt{N}l_{s}$ is important here. The
background possesses an asymptotically flat region $U\gg \frac{\sqrt{N}l_{s}}{l_p^3}$
connected to a semi-infinite flat tube $\frac{1}{l_{s}^{2}}\ll U\ll \frac{\sqrt{N}l_{s}}{l_p^3}$,
with the topology of $\mathbb{R}^{+}\times S^{3}\times \mathbb{R}^{6}$.
In this case, rather than a SCFT, the limiting theory on the brane will
be little string theory \cite{9808149}.

We now turn our attention to the pp-wave limit of the above geometry. 
The near-horizon limit of the NS5-brane metric \eqref{NS5} can be written in the
linear dilaton form
\begin{equation}
ds_{str}^{2}=Nl_{s}^{2}\left( -d\tilde{t}^{2}+\cos ^{2}\theta d\psi
^{2}+d\theta ^{2}+\sin ^{2}\theta d\phi ^{2}+\frac{dU^{2}}{U^{2}}\right)
+dx_{\mu }^{2}
,\end{equation}
where $x_{6}=\left( t,x_{\mu }\right)$, $\mu=1,\dots ,5$, and $t=\sqrt{N}l_{s}\tilde{t}$. 
Introducing new coordinates and keeping $\phi $ and $x_{\mu }$ unchanged,
\begin{equation}
  \tilde{t} + \psi =u,  \quad  \tilde{t} - \psi = 2\frac{v}{Nl_{s}^{2}},  \quad
  U =\frac{\sqrt{N} l_s}{l_p^3} \exp(y/\sqrt{N}l_{s}), \quad \theta =\frac{x}{\sqrt{N}l_{s}}
.\end{equation}
By taking the large-$N$ limit and keeping the rescaled coordinates fixed, \cite{0205258}
\begin{eqnarray}
ds_{str}^{2} &=&-2dudv-\frac{x_{i}^{2}}{4}du^{2}+dx_{i}^{2}+dx_{A}^{2},
\label{pp} \\
B_{ij} &=&u\epsilon _{ij},  \label{ppB}
\end{eqnarray}
where $x_{A}=\left(y,x_{\mu}\right)$ and $dx_{i}^{2}=dx^{2}+x^{2}d\phi^{2}$. 
This is the Nappi-Witten background \cite{9310112} with time-dependent $B$-field.
The gauge transformation
\begin{equation}
B_{\mu \nu }\rightarrow B_{\mu \nu }+\partial _{\mu }\lambda _{\nu}-\partial _{\nu }\lambda _{\mu }
\end{equation}
with $\lambda _{i}=\frac{1}{2}u\epsilon _{ij}z^{j}$ takes the $B$-field to 
\begin{equation}
B_{iu}=-\frac{1}{2}\epsilon _{ij}x^{j}.  \label{ppB-gauge}
\end{equation}
We can relate the light-cone energy and momentum to the original energies
and momenta as measured in the linear dilaton metric \cite{0205258}.
Recalling that $\partial _{t}=\frac{1}{\sqrt{N}l_s}\partial _{\tilde{t}}$, 
\begin{eqnarray}
\label{pmpp}
2p^{-} &=&-i\frac{\partial }{\partial u}=-i\left( \frac{\partial }
{\partial \tilde{t}}+\frac{\partial }{\partial \psi }\right) =\tilde{E}-J=
\sqrt{N}l_sE-J, \nn\\
2p^{+} &=&-i\frac{\partial }{\partial v}=-\frac{i}{Nl_s^2}\left( 
\frac{\partial }{\partial \tilde{t}}-\frac{\partial }{\partial \psi }\right) =
\frac{\tilde{E}+J}{Nl_s^2}=\frac{\sqrt{N}l_sE+J}{Nl_s^2}.
\end{eqnarray}
We will study string theory in the $N\rightarrow \infty $ limit, with $E\sim \sqrt{N
}$, $J\sim N$, while keeping $\sqrt{N}l_sE-J$ finite.

\section{$\sigma$-model and spectrum}
\label{smas}

In this section we apply light-cone quantisation to the NSR string in the
Nappi-Witten background. 
 The worldsheet fields in the light-cone gauge are as follows. 
The bosonic fields are $x^i$ with $i=1,2$, $x^A$ with $A=1...6$, $u$ and $v$.
The corresponding fermionic fields, which are real Majorana worldsheet spinors, are
$\psi^i$, $\psi^A$, $\psi^u$ and $\psi^v$. We use the conventions 
$\rho^\sigma= {{0 \ \  i} \choose {i \ \  0}}$, 
$\rho^\tau= {{0 \  -i} \choose {i \ \  0}}$, 
$\rho^3=\rho^\sigma\rho^\tau$,
$\eta^{\tau\tau}=-\eta^{\sigma\sigma}=-1$,
$\epsilon^{\tau\sigma}=+1$, and 
$\bar{\psi}=\psi^\dagger\rho^\tau$,
with the components of a spinor labeled as $\psi={\psi_- \choose \psi_+}$.

We begin with the NSR action in a curved background \cite{BRSSS,0006152},
\begin{eqnarray}
\label{NSRact}
2\pi\alpha'S=\int\td^2\sigma\sqrt{g}\bigg[
 	&-&\frac12 g^{\alpha\beta}\partial_\alpha x^\mu\partial_\beta x^\nu G_{\mu\nu}
	+\frac{i}2\bar{\psi}^\mu\rho^\alpha D_\alpha \psi^\nu G_{\mu\nu}
	\nn\\
	&-&\frac1{12}R_{\mu\nu\rho\sigma}\bar{\psi}^\mu\psi^\nu\bar{\psi}^\rho\psi^\sigma
	+\frac{k}{8\pi}\frac1{\sqrt{g}}\epsilon^{\alpha\beta}\partial_\alpha 
		x^\mu\partial_\beta x^\nu b_{\mu\nu}
	\nn\\
	&-&\frac{ik}{16\pi}\bar{\psi}^\mu\rho^\alpha\rho_3\psi^\nu\partial_\alpha 
		x^\lambda T_{\mu\nu\lambda}
\bigg]
,\end{eqnarray}
where $T=\td B$ is the field-strength for $B$, and $D$ denotes the covariant derivative
defined by
\begin{equation}
D_\alpha\psi^\mu=\partial_\alpha\psi^\mu+\Gamma_{\sigma\lambda}^\mu\partial_\alpha 
			x^\sigma\psi^\lambda
.\end{equation}
This action is worldsheet-supersymmetric for arbitrary $k$ and arbitrary background.
The existence of spacetime supersymmetry of course depends on the background, as well 
as the value of the parameter $k$. We will find in the Nappi-Witten background considered
next that we must have $k=4\pi$ to produce a supersymmetric spectrum.

The original Nappi-Witten background \cite{9310112} is specified by the metric
\begin{equation}
ds^2=G_{\mu\nu}dx^\mu dx^\nu={dx^A}^2+{dx^i}^2-2dudv+\bmfactor du^2
\end{equation}
and the $B$-field
\begin{equation}
B_{iu}=-\frac12 \epsilon_{ij}x^j
.\end{equation}
In the above metric, we shall be interested in the case $b=0$; it does no harm to consider $b\ne0$ for now,
and in fact the $b$-dependence will turn out to be inconsequential, merely adding a constant 
term to the Hamiltonian.
Substituting this background, the action \eqref{NSRact} becomes\footnote{We use the index notation $x^{i,A}$ to indicate that the summation is to be taken over values of both $i$ and $A$.}
\begin{eqnarray}
2\pi\alpha'S=\int\td^2\sigma\sqrt{g}\bigg[
	&-&\frac12 g^{\alpha\beta} \partial_\alpha x^{i,A}\partial_\beta x^{i,A}
	+g^{\alpha\beta}\partial_\alpha u \partial_\alpha v
	-\frac12 g^{\alpha\beta}\bmfactor 
		\partial_\alpha u \partial_\beta u 
	\nn\\
	&+&\frac{i}2 \bar{\psi}^{i,A}\rho\cdot\partial\psi^{i,A}
	+\frac{i}2 \frac{x^i}4 \bar{\psi}^i\rho\cdot\partial u\psi^u
	-\frac{i}2 \bar{\psi}^u\rho\cdot\partial\psi^v
	\nn\\
	&-&\frac{i}2 \frac{x^i}4 \bar{\psi}^u\rho\cdot\partial u \psi^i
	-\frac{i}2 \frac{x^i}4 \bar{\psi}^u\rho\cdot\partial x^i \psi^u
	-\frac{i}2 \bar{\psi}^v\rho\cdot\partial \psi^u
	\nn\\
	&-&\frac{i}2 \bmfactor \bar{\psi}^u\rho\cdot\partial \psi^u
	-\frac{1}{\sqrt{g}}\frac{k}{8\pi} \epsilon^{\alpha\beta}
		\partial_\alpha x^i \partial_\beta u \epsilon_{ij} x^j
	\nn\\
	&-&\frac{ik}{16\pi}\epsilon_{ij}\big(
		\bar{\psi}^i\rho^\alpha\rho_3\psi^j\partial_\alpha u
		+\bar{\psi}^j\rho^\alpha\rho_3\psi^u\partial_\alpha x^i
		+\bar{\psi}^u\rho^\alpha\rho_3\psi^i\partial_\alpha x^j
			\big)
\bigg]
.\end{eqnarray}
Varying this action with respect to $\bar{\psi}^v$ and $v$ leads to the 
equations of motion
\begin{equation}
	\rho\cdot\partial\psi^u=0
	\qquad {\rm{and}} \qquad
	\partial^2 u=0
\end{equation}
so, as in \cite{GSW} (pg.211) we may choose the light-cone gauge
\begin{equation}
\psi^u=0 \qquad , \qquad u=u_0+p^+\tau
,\end{equation}
where we have now made the choice $l_s=1$.
Imposing this gauge, and also $g_{\alpha\beta}=\eta_{\alpha\beta}$, 
we obtain the light-cone gauge action
\begin{eqnarray}
\label{LCGaction}
  \pi S_{\rm{LCG}}=\int\td^2\sigma\bigg[
	&-&\frac12 \eta^{\alpha\beta} \partial_\alpha x^{i,A}\partial_\beta x^{i,A}
	-p^+\partial_\tau v
	+\frac12\bmfactor {p^+}^2
	\nn\\
	&+&\frac{i}2 \bar{\psi}^{i,A} \rho\cdot\partial \psi^{i,A}
	+\frac{k}{8\pi}p^+\partial_\sigma x^i \epsilon_{ij} x^j
	-\frac{ik}{16\pi}p^+ \epsilon_{ij}\bar{\psi^i}\rho^\tau\rho_3\psi^j
  \bigg]
.\end{eqnarray}
This action may be transformed to a massless form when the constant $k$ is chosen to be $4\pi$.
We remark further on this point in section \ref{disc}. 

The canonical momenta conjugate to the $x$ and $\psi$ fields are
\begin{equation}
\pi^{i,A} = \frac{1}{\pi} \partial_\tau x^{i,A} 
\qquad , \qquad 
\xi^{i,A}=-\frac{i}{2\pi} \bar{\psi}^{i,A}\rho_\tau
.\end{equation}
The field $u$ is gauged away, while the term $p^+\partial_\tau v$ may be dropped as it is a total time-derivative.
The Hamiltonian is thus
\begin{eqnarray}
\label{Ham}
  p^+ H &=& -\frac{i}2 \bar{\psi}^{i,A}\rho_\sigma\partial_\sigma\psi^{i,A}
		  +\frac12(\partial_\tau x^{i,A})^2 + \frac12(\partial_\sigma x^{i,A})^2 
		  \nn\\
		&&  -\frac{kp^+}{8\pi}\left(
				\partial_\sigma x^i \epsilon_{ij} x^j 
                  		-\frac{i}2 \epsilon_{ij} \bar{\psi}^i\rho^\tau \rho_3 \psi^j 
				 \right)
		  -\frac{{p^+}^2}2 \bmfactor
.\end{eqnarray}

Varying the light-cone-gauge action \eqref{LCGaction} to obtain the equations of motion,
\begin{eqnarray}
	\delta x^A &\quad\rightarrow\quad& \partial^2 x^A=0
	,\\
	\delta x^i &\quad\rightarrow\quad& \big(\partial^2 - \frac{{p^+}^2}4\big)x_i
		=\frac{k}{4\pi}{p^+}\epsilon_{ij}\partial_\sigma x^j
	,\\
	\delta \bar{\psi}^A &\quad\rightarrow\quad& \rho\cdot\partial\psi^A=0
	,\\
	\delta \bar{\psi}^i &\quad\rightarrow\quad& \rho\cdot\partial\psi_i
		=\frac{k}{4\pi}{p^+}\epsilon_{ij}\rho^\tau\rho_3\psi^j
.\end{eqnarray}
The equations of motion for the $i$-labeled coordinates may be decoupled by defining 
$x^\pm=x^1\pm i x^2$ and $\psi^\pm = \psi^1 \pm i \psi^2$, giving
\begin{equation}
	\big(\partial^2 - \frac{{p^+}^2}4\big)x^\pm
		=\mp i \frac{kp^+}{4\pi} \partial_\sigma x^\pm
\end{equation}
and
\begin{equation}
	\rho\cdot\partial\psi^\pm
		=\mp i \frac{kp^+}{8\pi}\rho^\tau\rho_3\psi^\pm
.\end{equation}
Expanding in modes, we find that the frequencies for bosons and fermions match only when $k=4\pi$. 
Adopting this choice, we have
\begin{eqnarray}
	x^A=x_0^A+ p^A \tau + \frac{i}2 \sum_{n>0} 
	&\bigg(&
		\frac1{\sqrt{n}} a^A_n e^{-in(\tau-\sigma)} 
		 -\frac1{\sqrt{n}} {a^A_n}^\dagger e^{in(\tau-\sigma)}
		\nn\\
		&+&\frac1{\sqrt{n}} \tilde{a}^A_n e^{-in(\tau+\sigma)} 
		 -\frac1{\sqrt{n}} {\tilde{a}^A_n}{}^\dagger e^{in(\tau+\sigma)}
	\bigg)
\end{eqnarray}
with $x_0^A={x_0^A}^\dagger$ and $p^A={p^A}^\dagger$, and
\begin{eqnarray}
	x^+={x^-}^\dagger&=&\sqrt{\frac2{p^+}}{a_0^+}^\dagger e^{i\frac{p^+}2\tau}
		+\sqrt{\frac2{p^+}} a_0^- e^{-i\frac{p^+}2\tau}
		\nn\\
	   &+&e^{i\frac{p^+}2\tau}\sum_{n>0}\bigg(
		\frac{{a^+_n}^\dagger}{\sqrt{n+\frac{p^+}2}} e^{in(\sigma+\tau)}
		+\frac{{a^-_n}}{\sqrt{n-\frac{p^+}2}} e^{-in(\sigma+\tau)}
		\bigg)
		\nn\\
	   &+&e^{-i\frac{p^+}2\tau}\sum_{n>0}\bigg(
		\frac{{\tilde{a}^-_n}}{\sqrt{n+\frac{p^+}2}} e^{in(\sigma-\tau)}
		+\frac{{\tilde{a}^+_n}{}^\dagger}{\sqrt{n-\frac{p^+}2}} e^{-in(\sigma-\tau)}
		\bigg)
.\end{eqnarray}
In the fermionic sector, we have
\begin{eqnarray}
	\psi_-^A&=&d_0^A+\sum_{r>0}\bigg(
		d_r^A e^{-ir(\tau-\sigma)} + {d_r^A}^\dagger e^{ir(\tau-\sigma)}
	 \bigg) \\
	\psi_+^A&=&\tilde{d}_0^A+\sum_{r>0}\bigg(
		\tilde{d}_r^A e^{-ir(\tau+\sigma)} 
		+ {\tilde{d}_r^A}{}^\dagger e^{ir(\tau+\sigma)}
	 \bigg) 
\end{eqnarray}
and
\begin{eqnarray}
	e^{\pm i\frac{p^+}2\tau} \psi_-^\pm &=& 
		 d_0^\pm + \frac1{\sqrt{2}} \sum_{r>0} \big(
		   d_r^\pm e^{-ir(\tau-\sigma)} + {d_r^\mp}^\dagger e^{ir(\tau-\sigma)}
 		 \big) \\
	e^{\mp i\frac{p^+}2\tau} \psi_+^\pm &=& 
		 \tilde{d}_0^\pm + \frac1{\sqrt{2}} \sum_{r>0} \big(
		   \tilde{d}_r^\pm e^{-ir(\tau+\sigma)} 
                      + {\tilde{d}_r^\mp}{}^\dagger e^{ir(\tau+\sigma)}
 		 \big)
,\end{eqnarray}
where $d_0^A={d_0^A}^\dagger$ and $\tilde{d}_0^A={\tilde{d}_0^A}{}^\dagger$.
The summation in the fermionic sector may be over either integer or half-integer
values of $r$, corresponding to the Ramond and Neveu-Schwarz sectors; for 
half-integer values, the zero modes are to be omitted.

We impose canonical quantisation relations
\begin{equation}
\left[x^{i,A}(\sigma,\tau),\pi^{i',A'}(\sigma',\tau)\right]
= \left\{\psi^{i,A}_\pm(\sigma,\tau),\xi^{i',A'}_{\pm}(\sigma',\tau)\right\}
		= i\delta(\sigma-\sigma')\delta^{i,A\ i',A'}
,\end{equation}
which in terms of oscillators become:
\begin{equation}
\begin{array}{lcl}
[ x_0^A,p^B]=\delta^{AB}
&\qquad &
\\
{[} a_n^A , {a_m^B}^\dagger {]} = [\tilde{a}_n^A,{\tilde{a}_m^B}{}^\dagger]=\delta_{nm}\delta^{AB} 
&\qquad &
\{{d_r^A},{d_s^B}^\dagger\}=\{{\tilde{d}_r^A},{\tilde{d}_s^B}{}^\dagger\}=\delta_{rs}\delta^{AB}
\\
{[}a^\pm_n,{a^\pm_m}^\dagger ]=[\tilde{a}^\pm_n,{\tilde{a}^\pm_m}{}^\dagger]=\delta_{nm}
&\qquad &
\{d_r^\pm,{d_s^\pm}^\dagger\}=\{\tilde{d}_r^\pm,\tilde{d}_s^\pm{}^\dagger\}=\delta_{rs}
\end{array}
\end{equation}
The mode expansions and commutation relations for the bosonic fields agree
with those found in \cite{9503222} for the case of the bosonic string.

The Hamiltonian may now be written
\begin{eqnarray}
p^+ H &=&
 \frac{{p^A}^2}{2} + -b\frac{{p^+}^2}2 + \frac{p^+}2 ( {a_0^+}^\dagger a_0^+ +  {a_0^-}^\dagger{a_0^-} ) 
 - \frac{p^+}2 ({d_0^+}^\dagger d_0^+ +{\tilde{d}_0^+}{}^\dagger \tilde{d}_0^+)
 \nn\\
 && + \sum_{n>0} n({a_n^A}^\dagger a_n^A + {\tilde{a}_n^A}{}^\dagger\tilde{a}_n^A )
  +  \sum_{r>0} r({d_r^A}^\dagger d_r^A +{\tilde{d}_r^A}{}^\dagger \tilde{d}_r^A)
 \nn\\ 
 && + \sum_{n>0} \big(n+\frac{p^+}2\big)( {a^+_n}^\dagger a^+_n +  {\tilde{a}^-_n}{}^\dagger \tilde{a}^-_n)
  + \sum_{n>0} \big(n-\frac{p^+}2\big)( {\tilde{a}^+_n}{}^\dagger \tilde{a}^+_n +  {a^-_n}^\dagger a^-_n )
 \nn\\
 && +  \sum_{n>0} \big(n+\frac{p^+}2\big) ({d_r^+}^\dagger d_r^+ +{\tilde{d}_r^-}{}^\dagger \tilde{d}_r^-)
 +  \sum_{n>0} \big(n-\frac{p^+}2\big) ({d_r^-}^\dagger d_r^- +{\tilde{d}_r^+}{}^\dagger \tilde{d}_r^+)
,\end{eqnarray}
where $r=n$ or $r=n+1/2$ for the Ramond or Neveu-Schwarz sectors, respectively.
As dictated by supersymmetry, both the zero point energies and divergent terms cancel 
among bosonic and fermionic sectors.

Defining bosonic and fermionic number operators $N_n = a_n^\dagger a_n + \tilde{a}_n^\dagger \tilde{a}_n$ and 
$N_r = d_r^\dagger d_r + \tilde{d}_r^\dagger \tilde{d}_r$, we have as our final expression 
(setting $p^A=0$)
\begin{equation}
 H = \frac{1}{p^+} \sum_{n\ge0} \left( \sum_{A=1}^6 n (N_n^A + N_r^A)
+  (n+\frac{p^+}2) (N_n^+ + N_r^+ ) 
+  (n-\frac{p^+}2) (N_n^- + N_r^- ) \right)
,\end{equation}
where the same convention is used for the index $r$ as in the previous expression.
Supersymmetry is clearly evident in this expression; this is significant in that it 
demonstrates that the corresponding states of the dual LST must also fall into 
supersymmetric multiplets.

Writing this in terms of the original energy and momentum with \er{pmpp} we can
relate the energy to the angular momentum. Defining $\tilde{J}=J/\sqrt{N}$ we have
\begin{equation}
\label{ourBMN}
\bigg(E-\frac{\bar{N}}{\sqrt{N}}\bigg)^2
=\bigg(\tilde{J}+\frac{\bar{N}}{\sqrt{N}}\bigg)^2
+2\sum_n n \bigg[
	    \sum_{A=1}^6 (N_n^A+N_r^A) + N_n^+ + N_r^+ + N_n^- + N_r^- 
            \bigg]
\end{equation}
where
\begin{equation}
\bar{N} \equiv \frac14 \sum_n (N_n^+ + N_r^+ - N_n^- - N_r^-)
.\end{equation}
We emphasize here that $\bar{N}$ does not depend in the $A$-indexed modes.
We see from this expression that states for which $\bar{N}=0$ will exhibit
the standard relation between energy and angular momentum, while states with $\bar{N}\ne0$
shift $E$ and $J$, `trading' one for the other at order $1/\sqrt{N}$.

In the following section we shall compare this result with
the relation obtained via semi-classical analysis of a rotating
once-folded string \cite{0204226}.

\section{Semi-classical Analysis}

The semi-classical method allows for more general computations than
those done in the pp-wave limit \cite{0204051}.
In $AdS_{5}\times S^{5}$, the semi-classical method has been used
to first order, with a point-like string boosted in an $S^5$ direction,
to reproduce the result obtained via a pp-wave computation such
as in the previous section \cite{0204226,0209116}.

Here, following this example, we apply the semi-classical method 
to the case of the NS5-brane and compare with our pp-wave results.

We start with the linear dilaton metric
\begin{eqnarray}
ds_{str}^{2} &=&-dt^{2}+dx_{5}^{2}+\frac{Nl_{s}^{2}}{U^{2}}\left(
dU^{2}+U^{2}d\Omega _{3}^{2}\right)   \notag \\
&=&Nl_{s}^{2}\left( -d\tilde{t}^{2}+\frac{dU^2}{U^2}+d\Omega_3^2\right)
+d\rho^2+\rho^2d\Omega_4^2
\end{eqnarray}
where $t=\sqrt{N}l_{s}\tilde{t}$ and
\begin{eqnarray}
d\Omega_3^2&=&\cos^2 \theta d \psi^2 + d \theta^2 + \sin^2\theta d\phi^2 \\
d\Omega_4^2&=&d\beta_1^2+\cos^2\beta_1\left(d\beta_2^2+\cos^2\beta_2\left(d\beta_3^2+\cos^2\beta_3
d\beta_4^2\right)\right)
.\end{eqnarray}
Using the bosonic string action
\begin{equation}
\label{bsacg}
S=-\frac1{4\pi\alpha'}\int\td\sigma\td\tau G_{\mu\nu} \big( \partial_\sigma x^\mu \partial_\sigma x^\nu
	-\partial_\tau x^\mu \partial_\tau x^\nu \big)
,\end{equation}
with the above metric, the equation of motion may be obtained;
to find a classical solution, we make the ansatz
\begin{eqnarray}
\rho  &=&\rho \left( \sigma \right) ,U=U\left( \sigma \right) ,  \nn \\
\tilde{t} &=&\kappa \tau , \beta_4=\omega \tau ,\psi=\nu \tau ,  \nn \\
\theta&=&\phi=\beta_{1,2,3}=0,
\end{eqnarray}
which describes a string stretching along the directions $\rho$ and $U$,
rotating around the $\phi$ and $\varphi$ directions.
The action \eqref{bsacg} becomes
\begin{equation}
S=-\frac1{4\pi\alpha'}\int\td\sigma\td\tau \bigg[ N(\kappa^2-\nu^2)+\rho'^2-\omega^2\rho^2
	+ N\frac{U'^2}{U^2} \bigg]
,\end{equation}
and the resulting equations for $\rho $ and $U$ are
\begin{eqnarray}
\rho''+\omega ^{2}\rho  &=&0, \\
U''-\frac{U'^2}{U} &=&0,
\end{eqnarray}
with the constraint
\begin{equation}
T_{++}=T_{--}=-N\kappa^2+\rho'^2+\omega^2\rho^2+N\frac{U'^2}{U^2}+N\nu^2=0.
\end{equation}
The solution is
\begin{eqnarray}
\rho'^2&=&N(\kappa^2-\nu^2)\cos^2 \theta_0-\omega^2\rho^2 \\
U'^2&=&(\kappa^2-\nu^2) U^2 \sin^2 \theta_0
\end{eqnarray}
where $\theta_0$ is an integration constant.
 
We consider the once-folded string
configuration, with the string split into four segments; for $0<\sigma <\pi /2
$, the function $\rho \left( \sigma \right) $ increases from 0 to its
maximal value $\rho _0$. Then,
\begin{equation}
\rho ^{\prime }\left( \frac{\pi }{2}\right) =0 \quad \Rightarrow \quad \rho _0=
\frac{\sqrt{N(\kappa^2-\nu^2)\cos^2 \theta_0 }}{\omega }.
\end{equation}
Calculating the energy and angular momentum,
\begin{eqnarray}
\tilde{E} &=&-P_{\tilde{t}}=\frac{1}{4\pi \alpha ^{\prime }}\int_0^{2\pi
}\td\sigma 2N\kappa =\frac{N\kappa }{\alpha ^{\prime }}, \\
S &=&P_{\beta_4}=\frac{1}{4\pi \alpha ^{\prime }}\int_0^{2\pi }\td\sigma
 2\omega \rho ^{2}  \notag \\
&=&\frac{4\omega }{\pi \alpha ^{\prime }}\int_0^{\rho _0} \frac{
\rho ^{2}\td\rho }{\sqrt{N(\kappa^2-\nu^2)\cos^2 \theta_0 -\omega ^{2}\rho
^{2}}}=\frac{N(\kappa^2-\nu^2)\cos^2 \theta_0 }{2\alpha ^{\prime }\omega
^{2}}, \\
J &=&P_{\psi }=\frac{1}{4\pi \alpha ^{\prime }}\int_0^{2\pi }\td\sigma
 2N\nu =\frac{N\nu }{\alpha ^{\prime }}.
\end{eqnarray}
The relationship between $E$, $S$ and $J$ is thus
\begin{equation}
E^{2}=\frac{\tilde{E}^{2}}{N}=\frac{J^{2}}{N}+\frac{2\omega^2}{\alpha'\cos^2 \theta_0}S.
\end{equation}
Imposing the periodicity condition
\begin{equation}
2\pi =4\int_0^{\rho _0} 
\frac{\td\rho }{\sqrt{N(\kappa^2-\nu^2)\cos^2 \theta_0 -\omega ^{2}\rho ^{2}}}=\frac{2\pi }{\omega },
\end{equation}
we find $\omega=1$. Thus we obtain
\begin{equation}
\label{flatE}
E^{2}=\tilde{J}^{2}+\frac{2}{\alpha' \cos^2 \theta_0 } S 
.\end{equation}
In the case $\theta_0=0$ this expression for the energy of the string undergoing the motion
described above is the same as in the case of a flat background \cite{0204226}.

Identifying the order $n=1$ oscillator state with the spin $S$, as explained in ref. \cite{0204226},
we may apply our BMN formula \eqref{ourBMN} (in the bosonic sector) to obtain (reinstating $\alpha'$)
\begin{equation}
 \bigg(E-\frac{\bar{N}_1}{\sqrt{N}}\bigg)^2 = \bigg(\tilde{J}+\frac{\bar{N}_1}{\sqrt{N}}\bigg)^2 +\frac{2}{\alpha'}S
\end{equation}
where $\bar{N}_1$ refers to the $n=1$ contribution to $\bar{N}$.
In the large-$N$ limit this coincides with the relation \eqref{flatE}.
The reader is also referred to \cite{0302005,0002198}.
We conclude that the string theory in the pp-wave limit of the NS5-brane
background is dual to a sector of LST which has states in correspondence
to states of this free worldsheet theory.

\section{Discussion}
\label{disc}
In the Penrose limit, we are considering energy of order $\sqrt{N}l_{s}$ in
the sector with $J\sim N$. In the $AdS_{5}\times S^{5}$ case, the pp-wave light-cone worldsheet action
contains a `mass term'.
The mass term $x^2/2$ makes it different from string theory in
flat space-time. In the present case of an NS5-brane, the action \eqref{LCGaction}
contains both a mass term and antisymmetric B-field.
These two terms can be removed from our action by the following field redefinition
with the choice $k=4\pi$, which also is the choice which ensures space-time supersymmetry.
We redefine
\begin{eqnarray}
Z &=&e^{i\sigma p^+ /2}\left( x^{1}+ix^{2}\right) , \\
\Phi ^{+} &=&e^{i\sigma p^+ /2}\left( \psi ^{1}+i\psi ^{2}\right) , \\
\Phi ^{-} &=&e^{-i\sigma p^+ /2}\left( \psi ^{1}-i\psi ^{2}\right)
;\end{eqnarray}
the action becomes 
\begin{eqnarray}
S &=&\frac{1}{2\pi \alpha ^{\prime }}\int \td^2\sigma 
\sqrt{g}\left\{ -\frac{1}{2}g^{\alpha \beta }\left(
\partial _{\alpha }x^{A}\partial _{\beta }x^{A}+\partial _{\alpha }Z\partial
_{\beta }\bar{Z}-2\delta _{\alpha \tau }p^+\partial _{\beta }v\right) \right. 
\notag \\
&&\left. +\frac{i}{2}\bar{\psi}^{A}\rho ^{\alpha }\left( \partial _{\alpha
}\psi ^{A}\right) +\frac{i}{2}\left[ \frac{1}{2}\bar{\Phi}^{+}\rho
^{\alpha }\left( \partial _{\alpha }\Phi ^{+}\right) +\frac12\bar{\Phi}
^{-}\rho ^{\alpha }\left( \partial _{\alpha }\Phi ^{-}\right) \right]
\right\} ,
\end{eqnarray}
with the boundary conditions
\begin{eqnarray}
x^{A}\left( \sigma +2\pi \right)  &=&x^{A}\left(\sigma \right) ,\text{ \ }A=1,\dots ,6, \\
Z\left( \sigma +2\pi \right)  &=&e^{i\pi p^+}Z\left(\sigma \right) , \\
\Phi ^{+}\left( \sigma +2\pi \right)  &=&\pm e^{i\pi p^+}\Phi ^{+}\left( \sigma \right) , \\
\Phi ^{-}\left( \sigma +2\pi \right)  &=&\pm e^{-i\pi p^+}\Phi ^{-}\left( \sigma \right)
.\end{eqnarray}
We end up with a free worldsheet theory very similar to that of NSR
strings in flat spacetime. The difference in this case is that two of
the fields have twisted boundary conditions.  This is, of course, what
we found in the direct calculation of previous sections.

\section*{Acknowledgements}

The authors wish to thank R.~Rashkov for useful comments. This work
was supported in part by a grant from the Natural Sciences and
Engineering Research Council of Canada. R.~P. is grateful to the
Department of Physics at Simon Fraser University for kind hospitality.

\end{document}